\documentclass{scrartcl}
\usepackage{authblk}
\usepackage[numbers, sort]{natbib}
\usepackage{microtype, textcomp}
\usepackage[abbreviations=false, detect-all, list-units=single, separate-uncertainty=true]{siunitx}
\usepackage{amsmath, physics}
\usepackage[english]{babel}
\usepackage[utf8]{inputenc}
\usepackage{xspace}
\usepackage{graphicx}
\usepackage[section]{placeins}

\usepackage[final]{changes}
\colorlet{Changes@Color}{red}

\newcommand{\cypho}{\textit{Cyphochilus}\xspace}
\newcommand{\meep}{MEEP\xspace}
\newcommand{\mcpp}{MCPlusPlus\xspace}
\newcommand\CoAuthorMark{\footnotemark[\arabic{footnote}]}

\usepackage{hyperref}
\begin{document}

\title{Optimized white reflectance in photonic network structures}

\author[1]{Francesco Utel\footnote{These two authors contributed equally.}}

\author[2]{Lorenzo Cortese\protect\CoAuthorMark\thanks{lorenzo.cortese@icfo.eu; Corresponding author}}

\author[1,3]{Diederik S. Wiersma}

\author[1,3]{Lorenzo Pattelli\thanks{pattelli@lens.unifi.it; Corresponding author}}

\affil[1]{European Laboratory for Non-linear Spectroscopy (LENS), Università di Firenze, 50019 Sesto Fiorentino, Italy}
\affil[2]{Institut de Ciències Fotòniques (ICFO), The Barcelona Institute of Science and Technology, 08860 Castelldefels, Spain}
\affil[3]{Istituto Nazionale di Ricerca Metrologica (INRiM), 10135 Torino, Italy}

\date{\today}

\maketitle

{\footnotesize \sffamily\bfseries This is the peer reviewed version of the following article: Utel, F., Cortese, L., Wiersma, D. S., Pattelli, L., Optimized White Reflectance in Photonic‐Network Structures. \emph{Advanced Optical Materials} 2019, 7, 1900043, which has been published in final form at \url{https://doi.org/10.1002/adom.201900043}. This article may be used for non-commercial purposes in accordance with Wiley Terms and Conditions for Use of Self-Archived Versions.}

\begin{abstract}

\added{Three-dimensional disordered networks are receiving increasing attention as versatile architectures for highly scattering materials. However, due to their complex morphology, little is still known about the interplay between their structural and optical properties.}
\added{Here,} we \added{describe} a simple algorithm that allows to generate photonic network structures inspired by that of the \cypho beetle\added{, famous for the bright white reflectance of its thin cuticular scales}.
The model allows to vary the degree of structural anisotropy and filling fraction of the network independently, revealing the key contribution of these two parameters to the overall scattering efficiency.
\added{Rigorous} numerical simulations show that the obtained structures \added{can exceed the broadband reflectance} of the beetle while using less material, \replaced{providing}{which challenges previous claims of morphological optimization and provides} new insights for the design of advanced scattering materials.
\end{abstract}


Bright white materials owe their diffuse reflectance to the presence of multiple light scattering over a broad wavelength range \cite{wiersma2013disordered}.
Typically, ensuring that all visible frequencies undergo a large number of scattering events requires a combination of high refractive index materials and thick scattering layers.
In fact, there is a limit up to which scattering elements can be efficiently packed into a thin layer, above which the permittivity contrast drops due to effective medium considerations, and structural coloration effects can arise due to spatial correlations.
How these effects come into play when increasing the number of scattering elements is a complex topic which is under active investigation even for simple spherical scatterers \cite{naraghi2015near, aubry2017resonant, pattelli2018role}, with little results available for more realistic three dimensional disordered material distributions.

In particular, fibrous and network-like diffusers have recently raised considerable interest because of their outstanding scattering efficiency \added{\cite{burresi2014bright, pisignano2017perspectives, syurik2017bio, wilts2018evolutionary, zou2019biomimetic}}.
Depending on their orientation, elongated elements can exhibit \added{very} different scattering \added{cross sections}, and their collective alignment offers an additional degree of freedom to tune light transport properties in these media.
Indeed, shapes such as prolate ellipsoids or cylinders can be packed up to higher densities delaying the onset of spatial correlations at the cost of increased angular correlations \cite{bolhuis1997tracing, ferreiro2014random, meng2016maximally}.
Interestingly, both these aspects -- namely the high density and prevalent orientation exhibited by packed rods -- can contribute to increase the overall turbidity, which makes cylinders particularly suited to realize highly turbid materials with a flat response over a broadband wavelength range \cite{cortese2015anisotropic}.
Indeed, arguably the shortest transport mean free path reported to date in the visible range has been obtained in samples made of GaP nanocylinders \cite{muskens2009large} and, for low refractive index materials, in the chitinous network structure of the \cypho beetles \cite{burresi2014bright, cortese2015anisotropic}.
In recent years, the latter has inspired an array of bio-mimicking materials attempting to reproduce its outstanding efficiency in terms \added{of} strong light scattering and limited material usage, taking advantage of a wide range of fabrication techniques including electro-spinning, super critical CO\textsubscript{2} foaming, polymer phase separation and direct laser writing \added{\cite{haberko2013direct, cimadoro2018electrospun, syurik2017bio, pisignano2017perspectives, syurik2018bio, zou2019biomimetic}}, to name a few.

However, as opposed to nanoparticle systems, network materials are characterized by several additional aspects other than number density and spatial correlations, which makes it difficult to understand what key parameters should be optimized \added{to design highly scattering network structures}.
Few notable examples include phase percolation, angular correlation and network valence, all of which concur in determining their scattering properties \cite{liew2011photonic, imagawa2017robustness, cortese2015anisotropic, sellers2017local}.
In this respect, simple generative models for photonic structures allowing to investigate the effects of these parameters separately are much needed to gain insight on their role and relevance.

In this work, we describe a simple branching random walk (BRW) algorithm to generate random network structures inspired by that of the \cypho beetle. \added{Notably, the model} allows to control and vary independently the volume fraction and degree of angular correlations without altering structural parameters \added{such as the slab thickness, the shape and aspect ratio of the constituent elements}.
The optical and transport properties of the generated structures have been investigated through finite difference time domain (FDTD) calculations and an inverse Monte Carlo (MC) approach, showing \replaced{that the bright reflectance of the \cypho beetle can be easily matched and surpassed by acting solely on the degree of anisotropy and the volume fraction of the network. To the best of our knowledge, this is the first rigorous demonstration of the key role played by structural anisotropy in highly reflective disordered samples}{that it is possible to reproduce and even surpass the brightness of the \cypho beetle by considering just these two parameters, challenging previous claims on the optimized scattering efficiency of its structure}.

\begin{figure*}
\centering
\includegraphics[width=\linewidth]{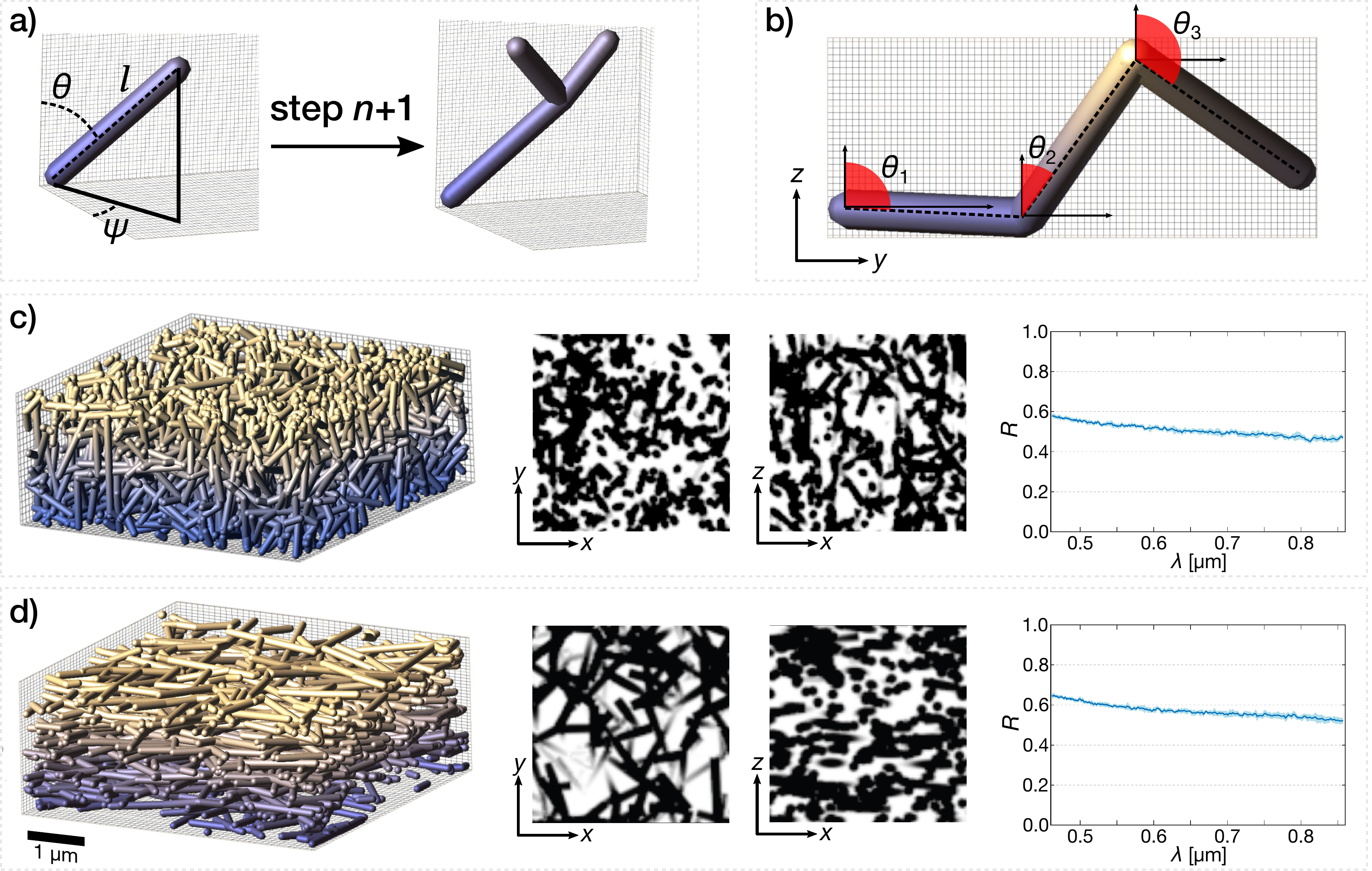}
\caption{Description of the BRW algorithm. a) At each iteration, two new rods are added on top of the previous endpoints. b) The polar angle $\theta$ is sampled from a normal distribution around $\mathrm{\pi}/2$ with a standard deviation $\sigma_\theta$, allowing to tune the degree of structural anisotropy (\added{panel} b). Two typical structures are shown in c)-d), obtained with a uniform and a narrow ($\sigma_\theta = \SI{0.2}{\radian}$) $\theta$ distribution, respectively. Tomographic sections show the prevalent in-plane alignment of the rods in the anisotropic structure. Both samples exhibit a broadband reflectance over the visible spectrum.}
\label{fig:BRW}
\end{figure*}

In our algorithm, a network structure is iteratively grown in a $7 \! \times \! 7 \! \times \! \SI{7}{\micro\meter\cubed}$ cubic cell with periodic boundary conditions (PBC) by adding rod-like scatterers corresponding to the steps of a three dimensional bifurcating random walk.
At each iteration, two new rods (steps) are attached to any loose end of the current network (walk), with length $l$ and direction $(\theta, \psi)$ drawn from predefined distributions (Figure\added{s} \ref{fig:BRW}a \added{and S1 in SI}).
\added{A starting condition of \num{\sim 1} rod/\SI{50}{\micro\meter\cubed} is set to obtain a more uniform material distribution.}
Network isotropy in the $xy$ plane is obtained by drawing the azimuthal angle $\psi$ uniformly between $\left[ 0, \mathrm{\pi}\right]$, while the the polar angle $\theta$ is drawn from a gaussian distribution centered around $\mathrm{\pi}/2$.
The degree of network anisotropy is tuned by changing the standard deviation $\sigma_\theta$ for the polar distribution, with smaller values corresponding to more layered/anisotropic arrangements.
Finally, an isotropic network is also generated for reference.
All rods have a fixed diameter of $d=\SI{250}{\nano\meter}$ and length $l \in (0,2\bar{l})$ sampled from a truncated gaussian distribution with mean $\bar{l}=\SI{1}{\micro\meter}$ and $\sigma_l = \SI{0.7}{\micro\meter}$ \added{in accordance with tomographic reconstruction of the beetle's structure \cite{vukusic2007brilliant, wilts2018evolutionary} (see also SI)}.
The refractive index of the rods is set to \num{1.56}, as in the case of chitin \added{\cite{leertouwer2011refractive, yoshioka2011direct}}.

Since our main focus is on the role of volume fraction and structural anisotropy, we \added{purposely minimize} the presence of spatial correlations by allowing each new rod to \added{intersect} previously placed elements. \added{This is not necessarily in contrast with the beetle structure, whose electron microscopy sections also show the occurrence of dense regions with touching or merged rods \cite{vukusic2007brilliant, luke2010structural}.}
To obtain more homogeneous material distributions, loose ends of newly placed rods falling closer than $2d$ to the extremity of any previously placed rod are \added{set to coincide,} and the growth of that branch is stopped.
The overall growth process stops when the target volume fraction $\varphi$ is reached.
Figure \ref{fig:BRW}c-d shows the 3D rendering of two illustrative portions of dense ($\varphi=\num{0.6}$) structures with isotropic or highly anisotropic alignment distributions, along with exemplary cross sections and their reflectance obtained by FDTD calculations.

Structures with different volume fractions of chitin and different degrees of structural anisotropy have been generated for each combination of $\varphi \in [ 0.1,\allowbreak 0.15,\allowbreak 0.2,\allowbreak 0.25,\allowbreak 0.3,\allowbreak 0.4,\allowbreak 0.5,\allowbreak 0.6,\allowbreak 0.7 ]$ and $\sigma_\theta \in [ 0.2,\allowbreak 0.5,\allowbreak 0.9,\allowbreak 1.2 ] \, \si{\radian}$, plus the isotropic case.
For the sake of simplicity, data corresponding to isotropic configurations are plotted at $\sigma_\theta = \mathrm{\pi}/2$.
The optical properties of the network are investigated using the FDTD software package \meep \cite{FDTDoskooi2010meep} on 3D cells with PBC along $x$ and $y$ and \SI{1.5}{\micro\meter}-thick perfectly matched layer (PML) along $z$ \replaced{(see SI)}{, for a final computation cell size of $7 \! \times \! 7 \! \times \! \SI{21.85}{\micro\meter\cubed}$}.
A plane-wave source emits a Gaussian pulse propagating along $z$ with central frequency $\nu_\text{c}=1/0.6$ in \meep units (corresponding to $\lambda_\text{c} = \SI{0.6}{\micro\meter}$) and full width $\Delta \nu = 1$.

\begin{figure*}
\centering
\includegraphics[width=\linewidth]{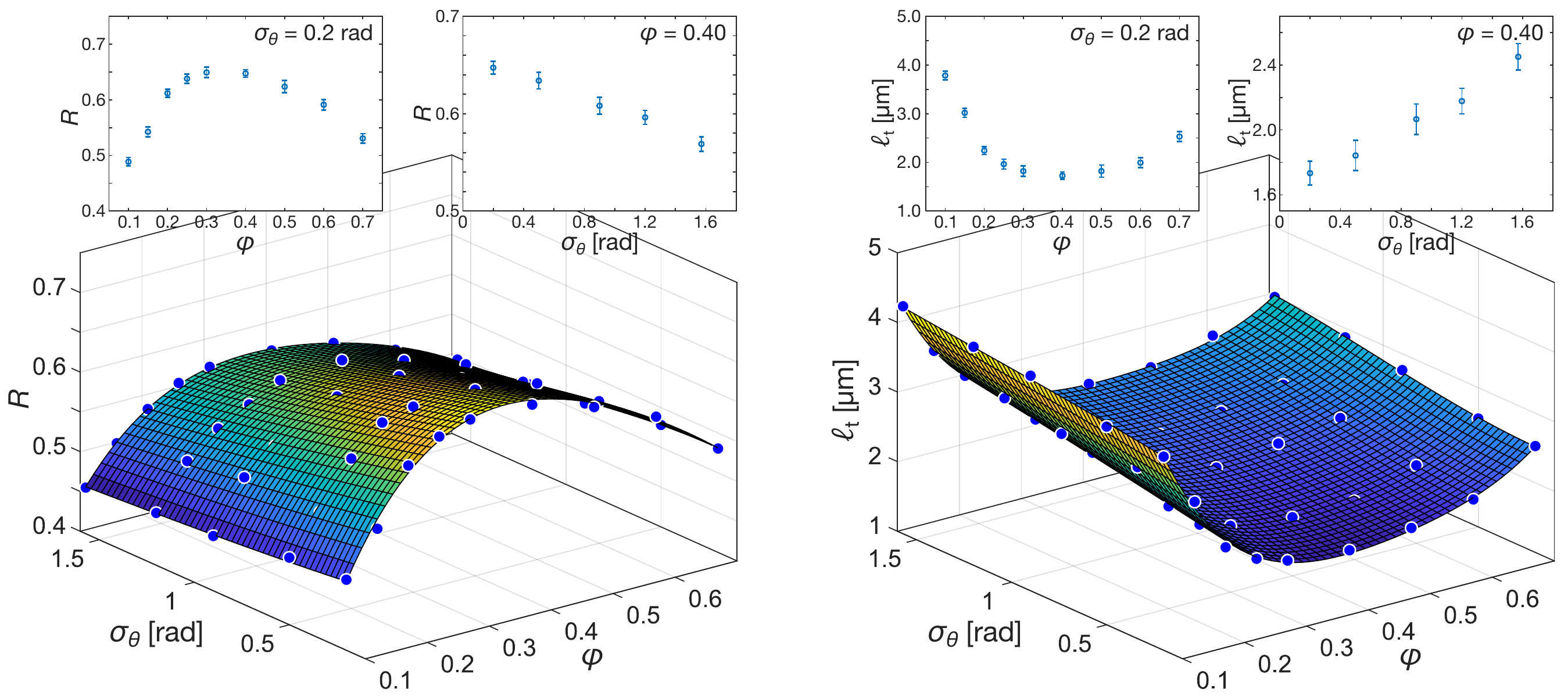}
\caption{Polynomial fitting of a) integrated reflectance at $\lambda = \SI{550}{\nano\meter}$ and b) retrieved transport mean free path $\ell_\text{t}$ along the perpendicular direction of generated structures as a function of BRW anisotropy $\sigma_\theta$ and final volume fraction $\varphi$. The isotropic case is displayed at $\sigma_\theta = \mathrm{\pi}/2$ for convenience. Insets show crosscuts at fixed anisotropy $\sigma_\theta = \SI{0.2}{\radian}$ or volume fraction $\varphi = \num{0.4}$. Each data point is averaged over 10 independent disorder realizations.}
\label{fig:surf}
\end{figure*}

All simulated structures within the explored parameter space exhibit a flat reflectance spectrum qualitatively similar to those shown in Figure \ref{fig:BRW}c-d, with varying brightness depending on the $(\varphi, \sigma_\theta)$ values.
Figure \ref{fig:surf}a shows an overview of the reflectance values obtained at $\lambda = \SI{550}{\nano\meter}$.
A maximum of the reflectance corresponding to optimized scattering efficiency is visible for highly anisotropic networks ($\sigma_\theta = \SI{0.2}{\radian}$) and intermediate volume fraction ($\num{0.3} < \varphi < \num{0.4}$).
Albeit lower, these values are in qualitative agreement with previous volume fraction measurements \cite{wilts2018evolutionary} and confirm that inducing a more `layered' orientation of the scattering rods provides a higher scattering efficiency along the perpendicular direction \cite{cortese2015anisotropic}.

To translate our results into a radiative transfer picture, we performed an inverse Monte Carlo analysis of the total reflectance data to retrieve corresponding transport mean free path values for each structure using the \mcpp software library \cite{mazzamuto2016deducing}.
\added{Compared to models based on the diffusion approximation, Monte Carlo inversion methods represent the gold standard to retrieve scattering parameters with high accuracy and precision \cite{hammer1995optical, dam2000multiple, palmer2006monte, hennessy2013monte}.}
We built a reflectance look-up table for each value of the volume fraction $\varphi$ and $\ell_\text{t} \in \num{0.1},\num{0.2},\dots,\SI{10}{\micro\meter}$ using a plane parallel slab geometry. \added{The effective refractive index $n(\varphi)$ is calculated using the Maxwell Garnett mixing formula for consistency with previous publications \cite{burresi2014bright}. Due to its symmetry properties, Bruggeman's formula represents another common model especially when the different materials occupy a comparable volume fraction \cite{sihvola1999electromagnetic}. We note, however, that due to the low permittivity contrast in our samples, results obtained with either approach are basically indistinguishable (see SI).}
Interpolating the look-up table allows to invert it providing an estimate of $\ell_\text{t}$ along $z$ for each configuration, as shown in Figure \ref{fig:surf}b.
A polynomial fit shows that the shortest transport mean free path is obtained for the most anisotropic structure tested ($\sigma_\theta = \SI{0.2}{\radian}$) and a filling fraction $\varphi = 0.39$.
At a wavelength of \SI{550}{\nano\meter}, the minimum value is $\ell_\text{t} = \SI{1.69}{\micro\meter}$, in qualitative agreement with previous experimental measurements \cite{burresi2014bright, cortese2015anisotropic}.

\begin{figure*}
\centering
\includegraphics[width=\linewidth]{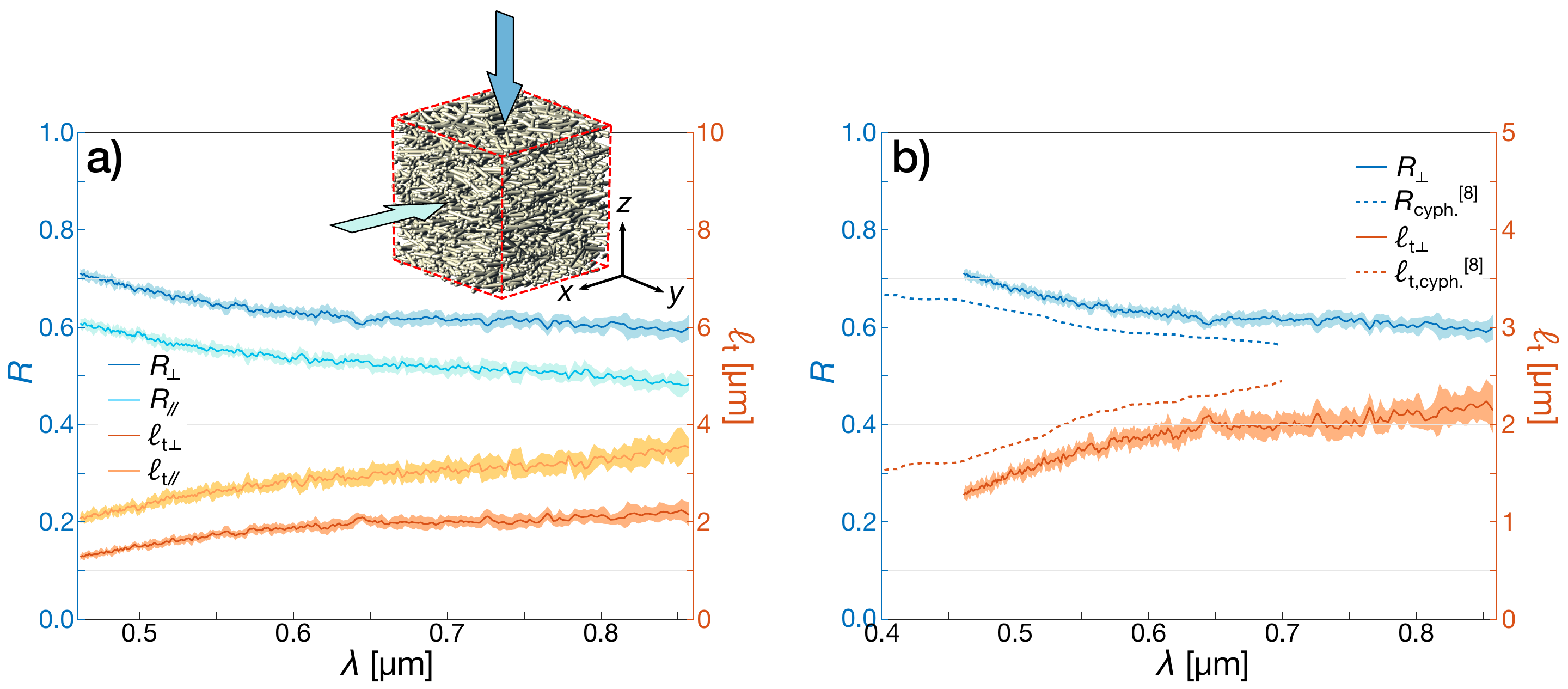}
\caption{Calculated reflectance of an optimally scattering sample with $\sigma_\theta = \SI{0.2}{\radian}$ and $\varphi = \num{0.4}$ a) illuminated at perpendicular and parallel incidence, with corresponding $\ell_\text{t}$ values retrieved from MC inversion and b) compared to the reflectance values obtained by FDTD calculations on the tomographic reconstruction of the \cypho ultrastructure of the same size \cite{wilts2018evolutionary}, showing a consistently higher reflectance in the visible range.}
\label{fig:ltbest}
\end{figure*}

Using the same method, it is possible to estimate the degree of transport anisotropy in the BRW structure by performing multiple FDTD calculations impinging on the cubic cell from different directions.
Figure \ref{fig:ltbest}a shows the results obtained on strongly scattering configurations ($\sigma_\theta = \SI{0.2}{\radian}$, $\varphi = \num{0.4}$) for light impinging at perpendicular (along $z$) and parallel (along $x$) incidence, which further confirms the higher diffuse reflectance of anisotropic structures when illuminated from the perpendicular direction to the plane of preferential alignment of the scattering elements.
Retrieved $\ell_\text{t}$ values for the two sample orientations show that the transport mean free path for lateral illumination is on average \num{1.63+-0.06} times longer than that at perpendicular incidence over the whole frequency range.

It is interesting to compare our results to those recently obtained by FDTD calculations on the tomographic 3D reconstruction of an actual \cypho scale \cite{wilts2018evolutionary} (Figure \ref{fig:ltbest}b).
In their work, Wilts \emph{et al.}\ have also studied the role of structural anisotropy and volume fraction by applying\replaced{, respectively, a uniaxial compression and an erosion/dilation transformation}{continuous deformations} to the tomographic reconstruction.
\added{However,} the concurrent modifications to \added{the slab thickness, its weight, the elementary rod shape and their aspect ratio induced by both these} deformations make it difficult to \replaced{assign unambiguously a reflectivity change to just one of these factors}{to draw conclusions regarding the optimization of the \cypho ultrastructure}.

Conversely, by tuning each parameter separately in our BRW generative algorithm, we are able to test their role without altering the \added{geometric} properties of the \added{elementary building blocks or the resulting slab assembly}.
We find that the diffuse reflectance obtained by an \added{optimized BRW} structure can exceed that of the actual \cypho beetle, resulting in a \SI{\sim16}{\percent} $\ell_\text{t}$ reduction along the perpendicular direction (Figure \ref{fig:ltbest}b)\added{ and a lower weight}.
In regard to previous claims on the optimally scattering performance of the \cypho structure \cite{burresi2014bright, wilts2018evolutionary}, our results show that either the degree of spatial or angular correlations (or both) in the elytral scales is sub-optimal, and that it is indeed possible to obtain significantly brighter structures improving the network morphology while also reducing its volume fraction.
Further optimization of both these aspects should be possible, given the naive and unrefined approach that we implemented for this simple algorithm.

It should be kept in mind that the chitinous network of the \cypho scales represents a trade-off between multiple goals, possibly including lightness, thermoregulation and mechanical stability.
Nonetheless, we note that our compenetrating BRW algorithm is also designed to form an interconnected phase through the unit cell to generate self-sustaining structures.
Moreover, we note that previous optimization claims are recently being further questioned by the demonstration of simple techniques allowing to fabricate \added{similar} network-like polymer structures with scattering properties that significantly outperform those of the \cypho beetle \added{\cite{syurik2018bio, zou2019biomimetic}}.

\added{In light of our findings, we argue} that the superior scattering efficiency exhibited by these simpler interconnected morphologies might be actually due to their \emph{less} uniform filament spacing, in contrast to the commonly invoked `optical crowding' effect. \added{In this respect, we see no apparent reason to envision the presence of fine-tuned spatial correlations to explain the optical turbidity of the beetle nor, consequently, to try to introduce them in the design of low-contrast network-like scattering materials -- largely relaxing fabrication accuracy requirements.}
Indeed, it has been recently shown that well-spaced scattering elements result in a lowered turbidity at volume fractions comparable or higher to that of the beetle \cite{pattelli2018role}.

It is worth stressing that the sub-optimal scattering efficiency of the \cypho structure is still very much consistent with an evolutionary picture, if not altogether expected.
From the evolutionary point of view, natural selection drives the inheritance of modified traits only up to when they can still provide an appreciable advantage in terms of reproduction success.
Beyond this point, there is no selective pressure to further maximize optical reflectance (or any other parameter) towards a relative optimum \added{\cite{fetzer2010evolution, johnson2013teaching}}.

In this respect, our simple BRW model provides a flexible platform to explore a larger configuration space, allowing to clarify the key features for efficient light scattering in network-like materials beyond the particular case of the \cypho beetle.
This stands in contrast with previously proposed models attempting to investigate its efficiency, which have been mainly limited to 2D projections, or involved ex-post deformations of the original structures rather than an ab-initio tuning of their growth parameters \cite{luke2010structural, wilts2018evolutionary, meiers2018bragg}.
Despite its simplicity, in fact, our BRW model retains a number of desirable features.
Firstly, as can be seen from Figure \ref{fig:BRW}, the final structures are still characterized by tomographic sections that are strikingly similar to those measured on the beetle \cite{vukusic2007brilliant}\added{, as also confirmed by their associated spectral density (see SI)}.
Secondly, by allowing rod intersections, we have obtained a model which reduces by construction any significant role of more elusive parameters such as spatial correlations, valence and local self-uniformity.
This largely simplifies the interpretation of our data, demonstrating that the bright reflectance from network-like materials is mainly due to the degree of their anisotropy and volume fraction\added{, guiding towards a more efficient design and fabrication of novel coatings}.
Nonetheless, the generative algorithm can be straightforwardly expanded to take also these additional aspects into consideration, e.g., enforcing a certain network valence at each node, using smoother rods and junctions, or controlling their degree of overlap through more refined algorithms\cite{altendorf2011random}.
Finally, BRW networks lend themselves naturally to FDTD calculations due to their periodic growth conditions, and can in principle be used in commercial direct laser writing fabrication processes by simply following the random walk growth process step by step.


In conclusion, we have designed a simple algorithm to generate network-like structures with tunable anisotropy for highly efficient, broadband light scattering.
Our algorithm takes inspiration from the complex network structure of the \cypho beetles scales and reduces it to just two basic properties, namely its volume fraction and degree of anisotropy.
To our knowledge, this allowed us to \replaced{unambiguously demonstrate and quantify the importance of structural anisotropy in random network materials for the first time.}{perform the first rigorous investigation of their role by varying these quantities independently and without altering the properties of the single scattering elements nor the sample thickness.}
Numerical calculations show that the obtained structures can match and even surpass the bright whiteness of an equivalent volume of the beetle's ultrastructure, demonstrating that additional factors such as spatial correlations or network valence do not play a significant role in explaining its high, broadband reflectance.

Our numerical results suggest that the elytral scales of the \cypho beetle do not represent an optimal structure in terms of weight and scattering efficiency, \replaced{as}{being arguably the result of a multi-objective optimization for optical, thermal and mechanical properties at the same time that did not necessarily reached a relative optimum.
In fact, we showed how one can easily obtain} similar morphologies \added{can be easily found} that are structurally stable, lighter and more turbid at the same time.
This general observation seems confirmed by the recent development of polymer-based materials which are clearly not as structurally refined as the beetle's chitinous network, and yet exhibit a significantly superior optical performance\added{ over the visible frequency range}.

Thanks to its simplicity and flexibility, we believe that our model can help deepen our understanding of the rich optical properties of interconnected network materials, as well as streamlining the generation, simulation and fabrication of novel materials with advanced scattering properties.

\section*{Acknowledgments}

We thank Matteo Burresi, \added{Fabrizio Martelli, }Silvia Vignolini and Bodo Wilts for fruitful discussion.
We acknowledge funding from the European Research Council under the European Union's Seventh Framework Program (No.~FP7/2007-2013)/ERC Grant Agreement No.~291349, Ente Cassa di Risparmio Firenze, Prog.~No.~2015-0781 and Laserlab-Europe, H2020 EC-GA (654148).

\section*{Conflict of interest}
The authors declare no conflict of interest.

\section*{Keywords}
Multiple scattering, bio-inspired photonic network, anisotropic light transport, light scattering optimization, scattering efficiency

\end{document}